\documentclass[aps,pra,twocolumn,superscriptaddress]{revtex4}
\usepackage{amsfonts,amssymb,graphicx}
\usepackage{epstopdf}
\usepackage{amsmath}
\usepackage[usenames,dvipsnames]{xcolor}
\usepackage{stackrel}
\usepackage[normalem]{ulem}
\usepackage{color}

\begin{document}

\title{Evolution of two-magnon bound states in a higher-spin ferromagnetic chain with single-ion anisotropy: A complete solution}
\author{Xinlan Lou\footnote{These two authors equally contributed to the work.}}
\affiliation{Center for Quantum Technology Research, School of Physics, Beijing Institute of Technology, Beijing 100081, China}
\affiliation{Key Laboratory of Advanced Optoelectronic Quantum Architecture and Measurements (MOE), School of Physics, Beijing Institute of Technology, Beijing 100081, China}
\author{Jiawei Li$^*$}
\affiliation{Center for Quantum Technology Research, School of Physics, Beijing Institute of Technology, Beijing 100081, China}
\affiliation{Key Laboratory of Advanced Optoelectronic Quantum Architecture and Measurements (MOE), School of Physics, Beijing Institute of Technology, Beijing 100081, China}
\author{Ning Wu}
\email{wunwyz@gmail.com}
\affiliation{Center for Quantum Technology Research, School of Physics, Beijing Institute of Technology, Beijing 100081, China}
\affiliation{Key Laboratory of Advanced Optoelectronic Quantum Architecture and Measurements (MOE), School of Physics, Beijing Institute of Technology, Beijing 100081, China}

\begin{abstract}
Few-magnon bound states in quantum spin chains have been long studied and attracted much recent attentions. For a higher-spin ferromagnetic XXZ chain with single-ion anisotropy, several features regarding the evolution of the low-lying two-magnon bound states with varying wave number were observed in the literature. However, most of these observations are only qualitatively understood due to the lack of analytical tools. By combining a set of exact two-magnon Bloch states and a plane-wave ansatz, we achieve a complete solution of the two-magnon problem in such a system. We identify parameter regions that support different types of two-magnon bound states, with the boundaries defined by algebraic equations. We discover for the first time a narrow region in which two single-ion bound states coexist. We show that the phase diagrams for distinct wave numbers are similar to each other, which enables us to map the evolution of the bound states to the rectilinear movement of a representative point for given parameters in a rescaled phase diagram. This dynamic picture provides quantitative interpretations of the observed features.
\end{abstract}

\maketitle
\par \emph{Introduction}. Recently, there has been a resurgence in the study of few-magnon bound states (BSs) due to the experimental observation of two-magnon BSs in the spin-1/2 ferromagnetic XXZ chain~\cite{Nature2013}, as first predicted by Bethe~\cite{Bethe1931}. Theoretically, the generalization of the two-magnon problem to higher spins and higher spatial dimensions was initiated by Dyson~\cite{Dyson1956}, followed by Wortis who found using Green's functions that the exchange (Ex) two-magnon BSs may exist in certain regions of the Brillouin zone~\cite{Wortis1963}. Silberglitt and Torrance~\cite{Torrance1970}, and Tonegawa~\cite{Tonegawa1970} made further generalizations to take the effect of single-ion (SI) anisotropy into account and revealed the existence of the so-called SI BSs. Papanicolaou and Psaltakis later extended Bethe's ansatz to the case of higher spins and solved the two-magnon problem in an XXX chain with SI anisotropy~\cite{Papanicolaou1987}.
\par In the case of ferromagnetic chains, it is well known that the two types of two-magnon BSs may appear as the lowest two branches of the excitation spectra. These two branches exhibit several remarkable features, including possible transitions between different types of BSs, the appearance of an alternative type of BS at certain wave numbers, and the existence of a branch consisting of a unique type of BS under certain parameters. Unfortunately, most of the understanding of these properties in the existing literature remains mainly qualitative due to the lack of appropriate analytical methods~\cite{Torrance1970,Papanicolaou1987,PRB2022}.
\par  In this work, we succeed in providing analytical and quantitative interpretations to the abovementioned properties. We develop a semianalytical method that combines a set of exact two-magnon Bloch states~\cite{PRB2022} and a plane-wave ansatz for solving a class of inhomogeneous tridiagonal matrices. For each wave number, we obtain a phase diagram that contains several phases supporting different types of bound states, with the phase boundaries determined by algebraic equations. We observe for the first time a narrow parameter region in which two SI BSs coexist. The phase diagrams for distinct wave numbers are shown to be similar to each other, which allows us to convert the evolution of BSs with varying wave number to the movement of a representative point for given parameters along a half-line in a rescaled phase diagram. This intriguing dynamic picture enables us to get analytical expressions for the wave numbers at which the foregoing transition or emergence take place.
\par \emph{Model and motivation}. We consider the spin-$S$ ferromagnetic XXZ ring in the presence of SI anisotropy~\cite{Torrance1970,Tonegawa1970,Papanicolaou1987,PRB2022}, which is described by the dimensionless Hamiltonian
\begin{eqnarray}\label{Haml}
\frac{H}{J}&=&-\sum^{N}_{j=1}[ (S^x_{j}S^x_{j+1}+S^y_{j}S^y_{j+1}+\Delta S^z_{j}S^z_{j+1})+d(S^z_{j})^2],\nonumber\\
\end{eqnarray}
where $S^{\alpha}_j$ are the spin operators on site $j$ with quantum number $S$, $J>0$ is the nearest-neighbor exchange interaction, $\Delta>0$ is the anisotropy parameter, and $d\geq 0$ measures the SI anisotropy strength. We assume that $N/2$ is even and $S>1/2$, and confine ourselves in the two-magnon subspace with total magnetization $NS-2$. Let $|F\rangle=|S,\ldots,S\rangle$ be the fully polarized state possessing an energy $E_F/J=-NS^2(\Delta+d)$, the real-space two-magnon basis states read $|i,j\rangle=\frac{1}{2S}S^-_iS^-_j|F\rangle$ ($1\leq i<j\leq N$) and $|i,i\rangle=\frac{1}{2\tilde{S}}(S^-_i)^2|F\rangle$ ($1\leq i \leq N$) with $\tilde{S}\equiv\sqrt{S(2S-1)}$. The two-magnon Bloch states are constructed as~\cite{PRB2022} $
|\xi_r(k)\rangle=\frac{e^{i\frac{rk}{2}}}{\sqrt{N}}\sum^{N-1}_{n=0}e^{ikn}T^{n}|1,1+r\rangle~(0\leq r<N/2)
$ with $k\in K_0=\left\{-\pi,-\pi+\frac{2\pi}{N},\ldots,0,\ldots,\pi-\frac{2\pi}{N}\right\}$, and $
|\xi_{\frac{N}{2}}(k)\rangle=e^{i\frac{Nk}{4}}\sqrt{\frac{2}{N}}\sum^{\frac{N}{2}-1}_{n=0}e^{ikn}T^{n}|1,1+\frac{N}{2}\rangle$ with $k\in K_1=\{-\pi,-\pi+\frac{4\pi}{N},\ldots,0,\ldots,\pi-\frac{4\pi}{N}\}$, where $T$ is the translation operator defined by $T|i,j\rangle=|i+1,j+1\rangle$. The complement of $K_1$ is denoted as $K'_1$ and we focus on $k\in K_1$ below. The matrix representation of $[H-E_F]/J$ in the ordered basis $\{|\xi_0(k)\rangle,|\xi_1(k)\rangle,\ldots,|\xi_{\frac{N}{2}}(k)\rangle\}$ is $
[\mathcal{H}_2(k)-E_F]/J=4 S \Delta+2(2S-1)d-h_2(k)$~\cite{PRB2022}, where
\begin{eqnarray}\label{h2}
h_2(k)=\left(
      \begin{array}{cccccccc}
       2d &   2\tilde{S}C_k &   & &    &    &   &   \\
          2\tilde{S}C_k & \Delta &  \tilde{C}_k &  &   &   &   &   \\
          &  \tilde{C}_k & 0 &  \tilde{C}_k  & &   &   &   \\
          &   &  \tilde{C}_k &  0 &  &    &   &   \\
           &   &    &  & \ddots&   &   &   \\
          &   &   &     & &0 &  \tilde{C}_k &   \\
          &   &   &    &  &\tilde{C}_k & 0 &   \sqrt{2}\tilde{C}_k \\
          &   &   &    & &  &    \sqrt{2}\tilde{C}_k & 0 \\
      \end{array}
    \right),\nonumber
\end{eqnarray}
is a $(N/2+1)\times (N/2+1)$ tridiagonal matrix with $C_k\equiv\cos\frac{k}{2}$ and $\tilde{C}_k\equiv 2SC_k$. % We thus converted the two-magnon problem to a single-particle one on an effective open chain with nonuniform onsite energies and hoppings.
Diagonalization of $[\mathcal{H}_2(k)-E_F]/J$ gives the excitation energies $\mathcal{E}^{(\alpha)}_{2}(k)$ and eigenstates $|\psi^{(\alpha)}(k)\rangle$ ($\alpha=1,2,\ldots,N/2+1$). We use the convention $
\mathcal{E}^{(1)}_{2}(k)\leq \mathcal{E}^{(2)}_{2}(k)\leq \ldots\leq \mathcal{E}^{(N/2+1)}_{2}(k)$ and write $
|\psi^{(\alpha)}(k)\rangle=\sum^{N/2}_{r=0}V^{(\alpha)}_r(k) |\xi_r(k)\rangle$. An eigenstate $|\psi^{(\alpha)}(k)\rangle$ is said to be an Ex (SI) BS if $
|V^{(\alpha)}_0(k)|<|V^{(\alpha)}_1(k)|>|V^{(\alpha)}_2(k)|>\ldots >|V^{(\alpha)}_{N/2}(k)|$ ($
|V^{(\alpha)}_0(k)|>|V^{(\alpha)}_1(k)|>|V^{(\alpha)}_2(k)|>\ldots >|V^{(\alpha)}_{N/2}(k)|$).
\begin{figure}
\includegraphics[width=.53\textwidth]{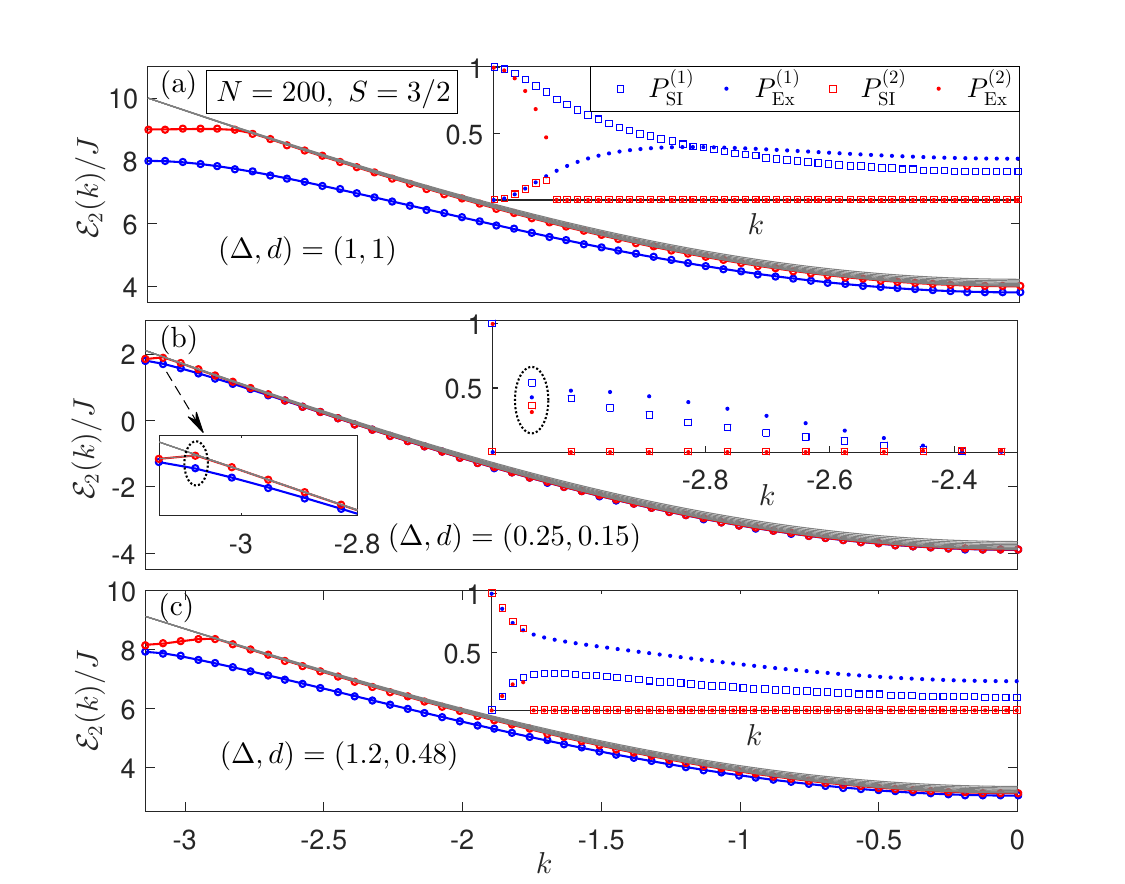}
\caption{The lowest 10 levels of the two-magnon excitation spectra calculated by numerically diagonalizing $[\mathcal{H}_2(k)-E_F]/J$ for $N=200,S=3/2$, and $k\in K_1$. The lowest two levels are indicated in colors; while the remaining eight ones form the narrow band (gray). The insets show the evolution of the weights of $|\xi_0(k)\rangle$ and $|\xi_1(k)\rangle$ in $|\psi^{(i)}(k)\rangle$ ($i=1,2$), $P^{(i)}_{\mathrm{SI}}(k)=|V^{(i)}_0(k)|^2$ (box) and $P^{(i)}_{\mathrm{Ex}}(k)=|V^{(i)}_1(k)|^2$ (dot).}
\label{motivation}
\end{figure}
\par We present in Fig.~\ref{motivation} several examples of the two-magnon spectra calculated by numerically diagonalizing $h_2(k)$ ($N=200$ and $S=3/2$). The insets show the weights of the two states $|\xi_0(k)\rangle$ and $|\xi_1(k)\rangle$, $P^{(i)}_{\mathrm{SI}}(k)=|V^{(i)}_0(k)|^2$ and $P^{(i)}_{\mathrm{Ex}}(k)=|V^{(i)}_1(k)|^2$ ($i=1,2$). For $(\Delta,d)=(1,1)$, the lowest separated branch (blue) was believed to be entirely contributed by the SI BSs~\cite{Torrance1970,Papanicolaou1987}. However, a closer look at $P^{(1)}_{\mathrm{SI/Ex}}(k)$ reveals that $|\psi^{(1)}(k)\rangle$ is actually an Ex BS for $-3\pi/5\leq k \leq 0$, as already been noticed by Tonegawa~\cite{Tonegawa1970}. The state $|\psi^{(2)}(k)\rangle$ (red) becomes an Ex BS at $k=-9\pi/10$. For $(\Delta,d)=(0.25,0.15)$, $|\psi^{(1)}(k)\rangle$ develops into an Ex BS for $-24\pi/25\leq k<-19\pi/25$, while $|\psi^{(2)}(k)\rangle$ remains a non-BS. Surprisingly, for the mode $k=-\pi+4\pi/N$ (circled by the dotted oval), both $|\psi^{(1)}(k)\rangle$ and $|\psi^{(2)}(k)\rangle$ are SI BSs. For $(\Delta,d)=(1.2,0.48)$, $|\psi^{(1)}(k)\rangle$ corresponds to Ex BSs throughout the zone and $|\psi^{(2)}(k)\rangle$ becomes an SI BS for $k\leq -47\pi/50$. The role of the two types of BSs is roughly interchanged compared with the case of $(\Delta,d)=(1,1)$.
\par These observations lead us to ask the following questions about the two lowest levels: How to analytically determine the transition point between different types of BSs and the wave number at which an alternative type of BS emerges? When will a unique type of BS occupy a single branch throughout the zone? Under what circumstance will two SI BSs simultaneously occur? Which parameter controls the interchange of the BSs? Except for Tonegawa's determination of the wave number at which a higher branch emerges~\cite{Tonegawa1970}, we did not find any other quantitative results concerning these questions.
\par
\begin{figure}
\includegraphics[width=.53\textwidth]{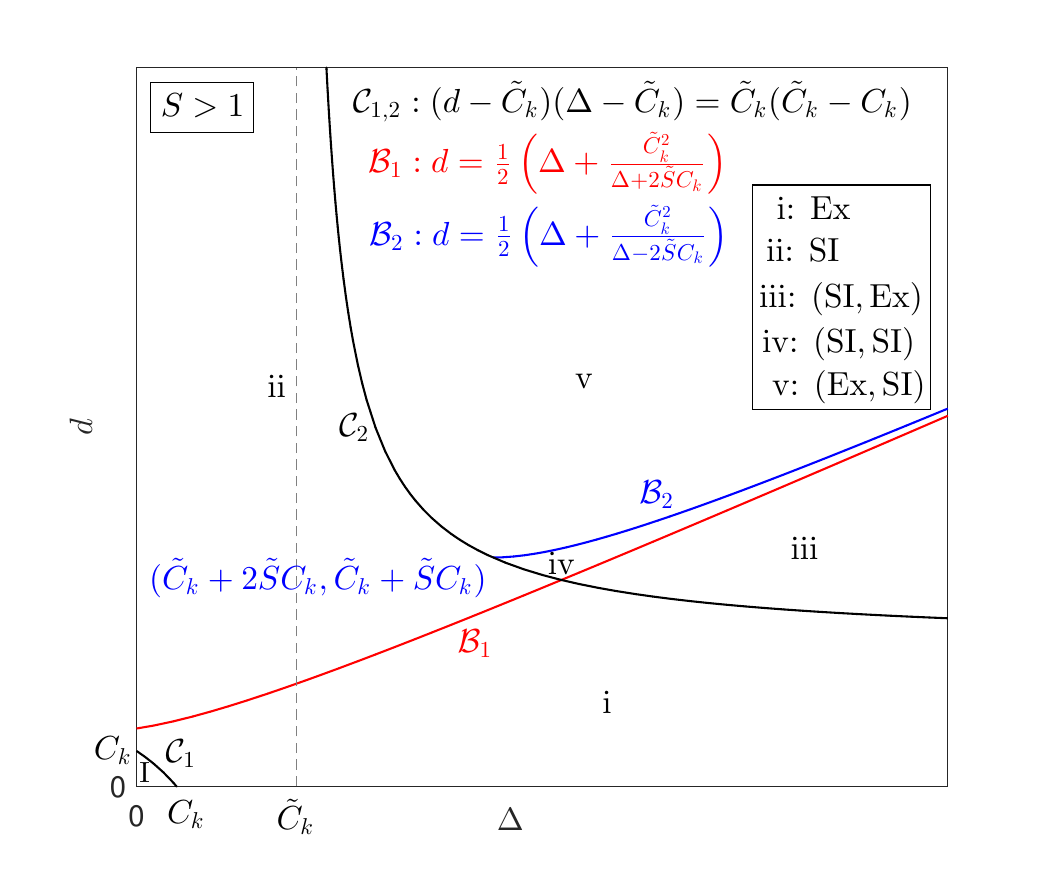}
\caption{For any $k\in K_1$, the curves $\mathcal{C}_{1,2}$ (black), $\mathcal{B}_1$ (red), and $\mathcal{B}_2$ (blue) divide the $\Delta-d$ plane into 6 regions I and i-v. Region i (ii) supports Ex (SI) BSs only. In region iii (v), the smaller and larger solution of Eq.~(\ref{Fy}) gives the SI and Ex (Ex and SI) BSs, respectively. In region iv we have two SI BSs.}
\label{phasediag}
\end{figure}

 \emph{Phase diagram: A plane-wave ansatz}. We now set out to provide a semianalytical solution of the problem and quantitatively answer the above questions. As detailed in \cite{SM}, the eigenvalue problem of $h_2(k)$ can be solved through a plane-wave ansatz $V_j=Xe^{ipj}+Ye^{-ipj}~(j=1,2,\ldots,N/2-1)$~\cite{Grimm,PRB2017,PRB2019,PRB2024}, where $X$ and $Y$ are $j$-independent coefficients. The `momentum' $p$ is determined by the solutions of the following transcendental equation,
\begin{eqnarray}\label{tan1}
 \tan \frac{N}{2}p&=& \frac{ w_k (\cos p)}{ (A_k-\cos p)\sin p},
\end{eqnarray}
where $A_k\equiv d/\tilde{C}_k- (\tilde{C}_k-C_k)/\Delta$ and $w_k(x)\equiv x^2-(A_k+\tilde{C}_k/\Delta)x+d/\Delta$. The number of real solutions of Eq.~(\ref{tan1}) is essentially determined by the sign of $w_k(1)=-[(d-\tilde{C}_k)(\Delta-\tilde{C}_k)-\tilde{C}_k(\tilde{C}_k-C_k)]/(\tilde{C}_k\Delta)$~\cite{SM}. The condition $w_k(1)=0$ defines the curves $\mathcal{C}_{1,2}$ as shown in Fig.~\ref{phasediag}. It can be shown that Eq.~(\ref{tan1}) has $N/2+1$, $N/2$, and $N/2-1$ real solutions in regions I, $\mathrm{i}\bigcup\mathrm{ii}$, and $\mathrm{iii}\bigcup\mathrm{iv}\bigcup\mathrm{v}$, respectively~\cite{SM}. For each real solution $p$, the excitation energy and the corresponding eigenstate are given by $\mathcal{E}_{2,p}(k)/J=4S\Delta+2(2S-1)d-2\tilde{C}_k\cos p$ and $V_{j,p}=\cos[(N/2+1-j)p]$ ($j=1,2\ldots,N/2-1$)~\cite{SM}. For large $N$, these excitation levels form the scattering continuum with oscillating wave functions.
\par In $\mathrm{i}\bigcup\mathrm{ii}$ and $\mathrm{iii}\bigcup\mathrm{iv}\bigcup\mathrm{v}$, Eq.~(\ref{tan1}) has purely imaginary solutions $p=i\tilde{p}$ (with $\tilde{p}$ real) satisfying~\cite{SM}
\begin{eqnarray}\label{tanh1_even}
\tanh (N\tilde{p}/2)&=& \theta(\cosh\tilde{p}),
\end{eqnarray}
where $\theta(x)\equiv -w(x)/[(A_k-x)\sqrt{x^2-1}]~(x>1)$. The excitation energy for each solution $\tilde{p}$ on $(0,\infty)$ is $\mathcal{E}_{2,\tilde{p}}(k)/J=4S\Delta+2(2S-1)d-2\tilde{C}_k\cosh \tilde{p}<\mathcal{E}_{2,p}(k)$, showing that the excitation level lies below the continuum. The wave functions of the associated BS read~\cite{SM}
\begin{eqnarray}\label{wf}
V_{j,\tilde{p}}&=&\cosh[(N/2+1-j)\tilde{p}]~(j=1,2,\ldots,N/2-1),\nonumber\\
V_{0,\tilde{p}}&=&RV_{1,\tilde{p}},\nonumber\\
V_{N/2,\tilde{p}}&=&V_{N/2-1,\tilde{p}}/(\sqrt{2}\cosh\tilde{p}),
\end{eqnarray}
where
\begin{eqnarray}\label{Rex}
R\equiv \tilde{S}C_k/(\tilde{C}_k\cosh\tilde{p}-d).
\end{eqnarray}
Since $V_{1,\tilde{p}}>V_{2,\tilde{p}}>\ldots>V_{N/2,\tilde{p}}$, the state is an SI (Ex) BS if $|R|>1$ ($0<|R|<1$). The boundaries in the $\Delta-d$ plane that separate the SI and Ex BSs, denoted by $\mathcal{B}_{1,2}$, are determined by the relations $R=\pm 1$. The solutions of Eq.~(\ref{tanh1_even}) on $\mathcal{B}_{1,2}$ are obviously $\tilde{p}_{\mathcal{B}_{1,2}}=\cosh^{-1}[d/\tilde{C}_k\pm\tilde{S}/(2S)]$.
\par In region $\mathrm{iii}\bigcup\mathrm{iv}\bigcup\mathrm{v}$ with $A_k>1$, there is a smaller solution of Eq.~(\ref{tanh1_even}), $x_{\mathrm{L}}$, on $(0,\cosh^{-1}A_k)$. The larger one $x_{\mathrm{R}}$ lies on $(\cosh^{-1}A_k,\infty)$~\cite{SM}. Clearly, $\tilde{p}=0$ is the unique (smaller) solution on $\mathcal{C}_1$ ($\mathcal{C}_2$). The minimal (maximal) value of $R$ on $\mathcal{C}_1$ is reached at $d=0$ ($d=C_k$): $R^{(\mathcal{C}_1)}_{\min}=\tilde{S}/(2S)<1$, $R^{(\mathcal{C}_1)}_{\max}=S/\tilde{S}\leq 1$. Thus, $\mathcal{C}_1$ never intersects $\mathcal{B}_{1}$ unless $S=1$; and $\mathcal{C}_2$ (with $R<0$) intersects $\mathcal{B}_{2}$ at $(\Delta,d)=(\tilde{C}_k+2\tilde{S}C_k,\tilde{C}_k+\tilde{S}C_k)$.
\par We emphasize that all the results obtained so far are exact for finite $N$. Moreover, our method combining the Bloch states and the plane-wave ansatz offers a simpler and more transparent treatment of the problem than the conventional Bethe ansatz~\cite{Papanicolaou1987,Vlad1984,Koma1997}, where the solutions of the associated Bethe ansatz equations could be subtle even for $S=1/2$~\cite{Vlad1984,Koma1997}.
\begin{figure}
\includegraphics[width=.53\textwidth]{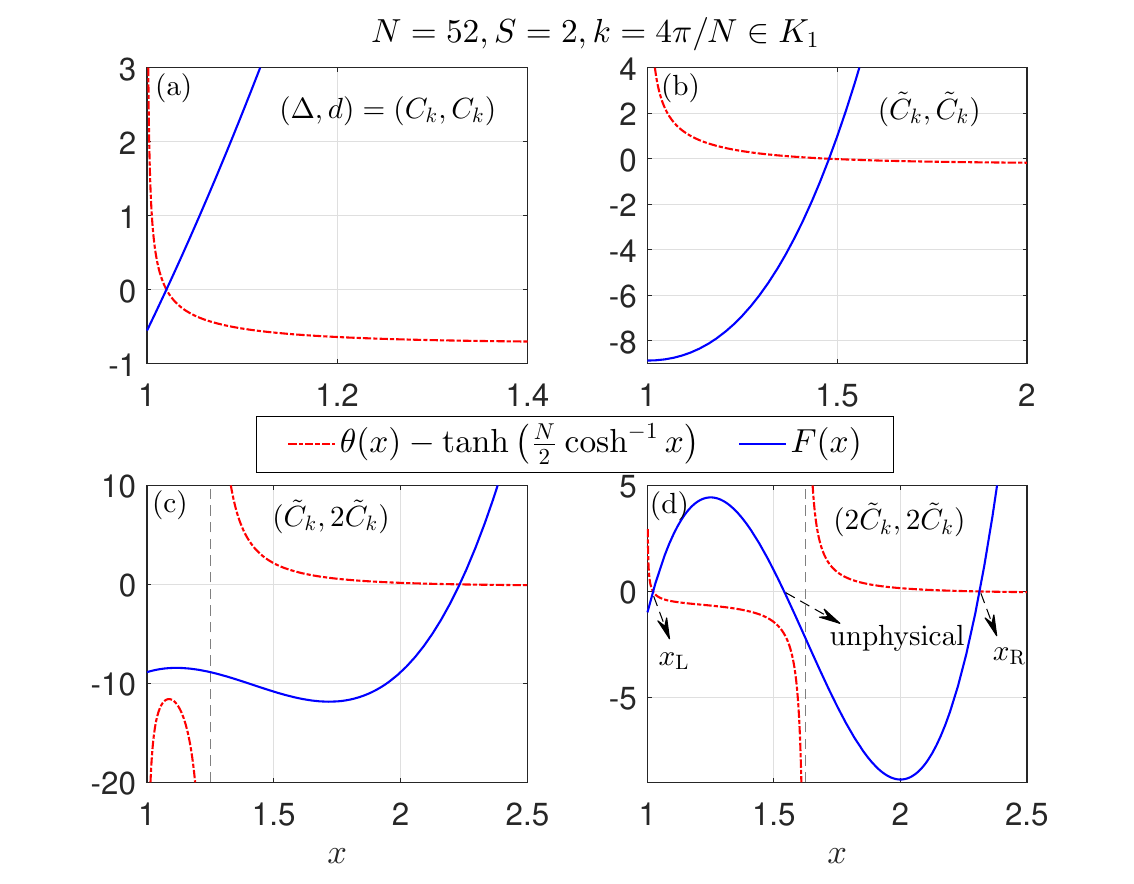}
\caption{Graphs of the functions $\theta(x)-\tanh\left(\frac{N}{2}\cosh^{-1}x\right)$ (red dot-dashed) and $F(x)$ [blue solid, Eq.~(\ref{Fx})] for $(\Delta,d)=(C_k,C_k)$, $(\tilde{C}_k,\tilde{C}_k)$, $(\tilde{C}_k,2\tilde{C}_k)$, and $(2\tilde{C}_k,2\tilde{C}_k)$. We choose $N=52$, $S=2$, and $k=4\pi/N\in K_1$. It can be seen that the solutions of Eq.~(\ref{tanh1_even}) are well approximated by the zeros of $F(x)$ given by Eq.~(\ref{Fx}). Note that for $(\Delta,d)=(2\tilde{C}_k,2\tilde{C}_k)\in \mathrm{iii}\bigcup\mathrm{iv}\bigcup\mathrm{v}$ there is an unphysical solution of Eq.~(\ref{Fy}) between the two physical solutions $x_{\mathrm{L}}$ and $x_{\mathrm{R}}$.}
\label{Fig1dot5}
\end{figure}
\par It is more desirable to get explicit expressions for the equations describing $\mathcal{B}_{1,2}$~\cite{whylargeN}. To this end, we consider the limit of large $N$ and regions with $\mathcal{C}_{1,2}$ excluded, where Eq.~(\ref{tanh1_even}) has no solutions near $\tilde{p}=0$. We can thus use the approximation $\tanh(N\tilde{p}/2)\approx 1$ for large $N$. Accordingly, Eq.~(\ref{tanh1_even}) is reduced to a cubic equation for $x\equiv \cosh\tilde{p}$,
\begin{eqnarray}\label{Fy}
F(x)&=&0,
\end{eqnarray}
where
\begin{eqnarray}\label{Fx}
F(x) &\equiv& 2\tilde{C}_k\Delta x^3-(\tilde{C}^2_k+2\tilde{C}_k A_k\Delta+2d\Delta+\Delta^2)x^2 \nonumber\\
&&+2(A_k d\Delta+A_k\Delta^2+\tilde{C}_kd)x-(d^2+A^2_k\Delta^2).\nonumber\\
\end{eqnarray}
As shown in Fig.~\ref{Fig1dot5}, for large $N$ (here $N=52$) the zeros of $F(x)$ give very accurate approximations to the actual solutions of Eq.~(\ref{tanh1_even}). By inserting $x_{\mathcal{B}_{1,2}}=\cosh\tilde{p}_{\mathcal{B}_{1,2}}$ into Eq.~(\ref{Fy}), we obtain the equations for $\mathcal{B}_{1,2}$:
\begin{eqnarray}\label{B11}
d=[\Delta+\tilde{C}^2_k/(\Delta\pm 2\tilde{S}C_k)]/2.
\end{eqnarray}
\par  Now, the curves $\mathcal{C}_{1,2}$ and $\mathcal{B}_{1,2}$ divide the $\Delta-d$ plane into six regions, I and i$-$v (Fig.~\ref{phasediag}). In region i (ii), we have only the Ex (SI) BS with $R<1$ ($R>1$). In region iii, the two solutions $x_{\mathrm{L}}$ and $x_{\mathrm{R}}$ correspond respectively to the SI ($R<-1$) and Ex ($R<1$) BSs, with the latter being the lowest-energy state. In region v, $x_{\mathrm{L}}$ and $x_{\mathrm{R}}$ give the Ex ($-1<R<0$) and SI ($R>1$) BSs, respectively. Interestingly, in the narrow region iv sandwiched by $\mathcal{C}_2,\mathcal{B}_1$, and $\mathcal{B}_2$, both of the two BSs are of SI-type (with $R<-1$ and $R>1$). This region is so narrow that it has most likely never been noticed in prior literature. Moreover, region iv would become narrower and narrower as $k$ shifts to the zone edge.
\par
\begin{figure}
\includegraphics[width=.53\textwidth]{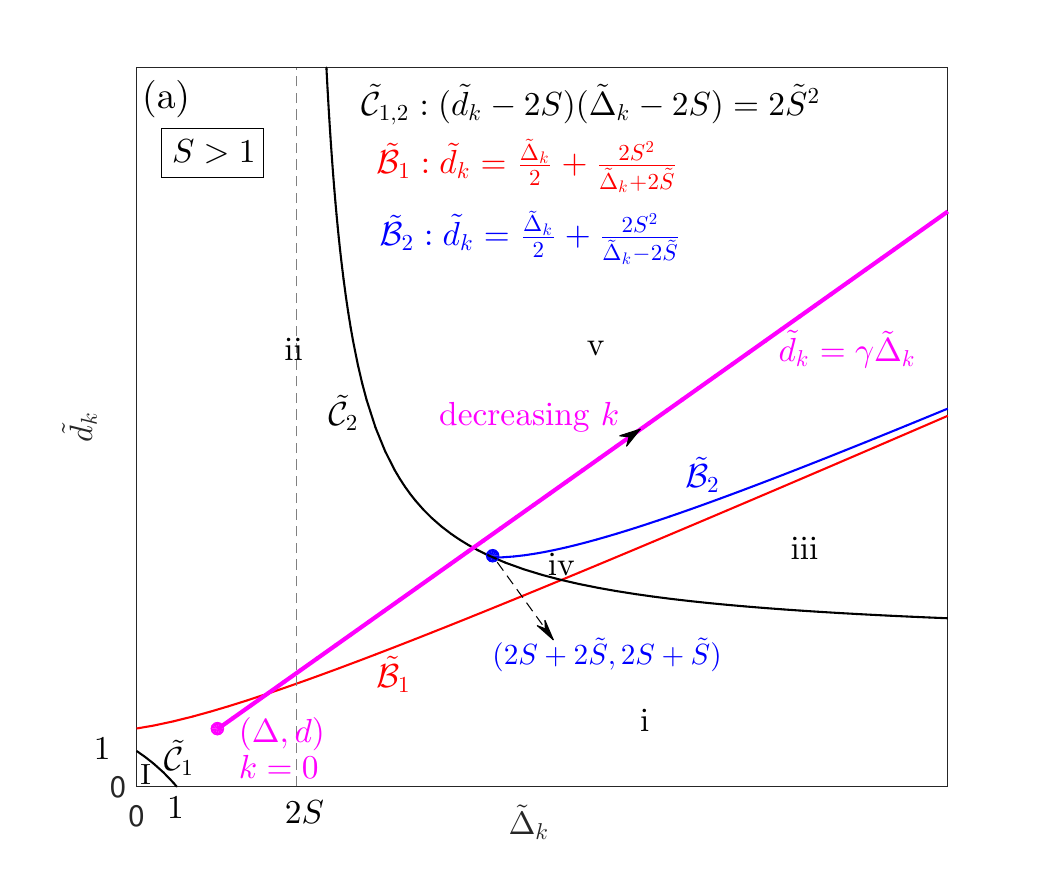}
\includegraphics[width=.53\textwidth]{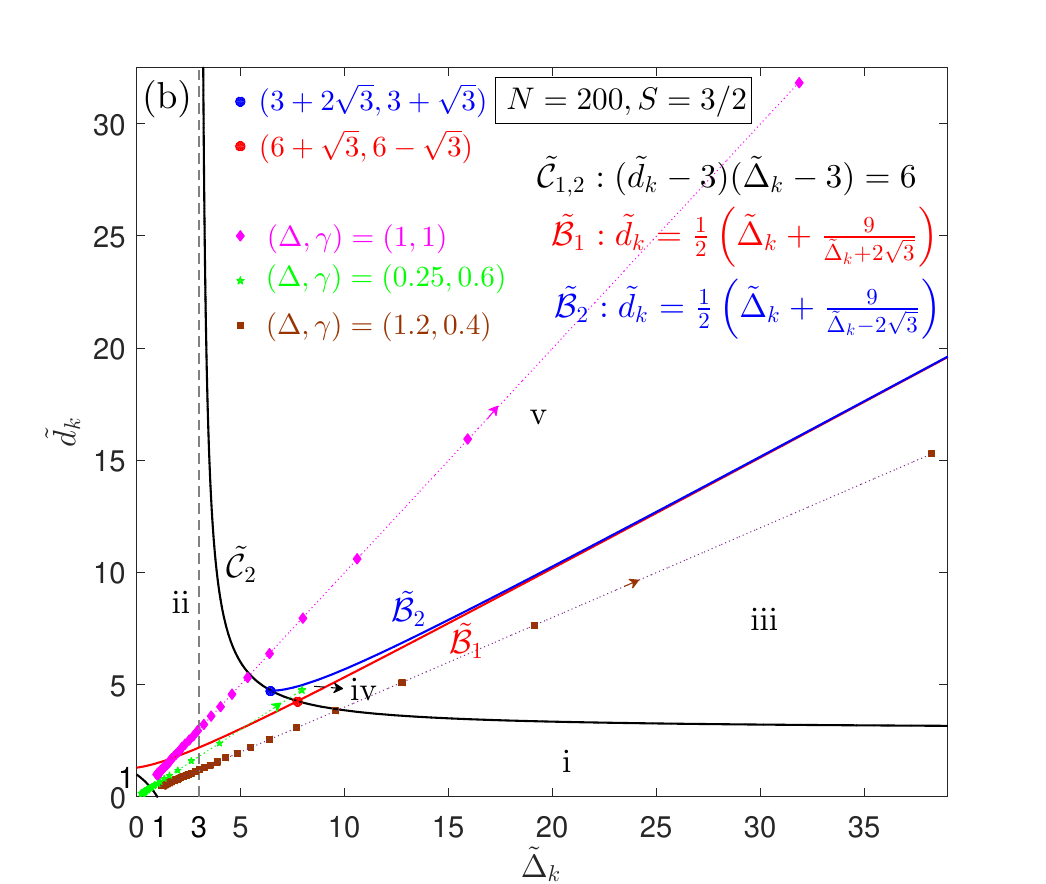}
\caption{(a) Rescaled phase diagram with coordinates $(\tilde{\Delta}_k,\tilde{d}_k)=(\Delta,d)/C_k$. The pink half-line $\tilde{d}_k=d\tilde{\Delta}_k/\Delta$ shows a typical trajectory of the representative points. (b) Evolution of the representative points for $N=200$, $S=3/2$, and $(\Delta,\gamma)=(1,1)$ (pink diamonds), $(0.25,0.6)$ (green stars), $(1.2,0.4)$ (brown squares). The parameters are chosen in accordance with those in Fig.~\ref{motivation}. The arrows indicate the direction of decreasing $k$ from $k=0$ to $k=-\pi+4\pi/N$ (note that representative point for $k=-\pi$ is at infinity).}
\label{mainresult}
\end{figure}

 \emph{Evolution of the BSs: A dynamic picture}. For any constant $\lambda$, the vertical line $\Delta=\lambda C_k$ intersects (if it does) with $\mathcal{C}_{1,2}$ and $\mathcal{B}_{1,2}$ at $d=[2S(\lambda-1)/(\lambda-2S)]C_k$ and $d=[\lambda/2+2S^2/(\lambda\pm2\tilde{S})]C_k$, indicating that the phase diagrams for distinct $k$'s are similar to each other. We are thus encouraged to set the $C_k$ for any $k$ as unity in the same figure. This procedure amounts to magnifying the phase diagram for $k\neq 0$ by a factor of $1/C_k$.  As a result, the variation of $k$ is mapped to the movement of the representative point (RP) for $(\Delta,d)$, $(\tilde{\Delta}_k,\tilde{d}_k)=(\Delta,d)/C_k$, in such a rescaled diagram. The equations for the borderlines $\mathcal{C}_{1,2}$ and $\mathcal{B}_{1,2}$ in the rescaled diagram become $\tilde{\mathcal{C}}_{1,2}:~(\tilde{d}_k-2S)(\tilde{\Delta}_k-2S)=2\tilde{S}^2$ and $\tilde{\mathcal{B}}_{1,2}:~\tilde{d}_k=\tilde{\Delta}_k/2+2S^2/(\tilde{\Delta}_k\pm2\tilde{S})$ [Fig.~\ref{mainresult}(a)].
\par The RP for $k=0$ is simply $(\tilde{\Delta}_0,\tilde{d}_0)=(\Delta,d)$; while the one for $k=-\pi$ is at infinity. As $k$ decreases, the RP will move along the half-line $\mathcal{L}_{\Delta,\gamma}:~\tilde{d}_k=\gamma \tilde{\Delta}_k$, with $\tilde{\Delta}_k\geq \Delta$ and $\gamma\equiv d/\Delta$. A typical example for the evolution of the RPs is illustrated by the pink half-line in Fig.~\ref{mainresult}(a), where $\mathcal{L}_{\Delta,\gamma}$ starts with $(\Delta,d)$ and extends to the upper-right direction as $k$ decreases. Clearly, the evolution of the BSs depends on two factors: the slope $\gamma$ and the value of $\Delta$. The former determines the regions through which the RPs can pass and the latter fixes the starting point. $\mathcal{L}_{\Delta,\gamma}$ may (if it does) intersect $\tilde{\mathcal{C}}_{1,2}$ and $\tilde{\mathcal{B}}_{1,2}$ at, respectively,
\begin{eqnarray}
\Delta_{\tilde{\mathcal{C}}_{1,2}}&=&[S(1+\gamma)\mp\sqrt{S^2(1+\gamma)^2-2\gamma S}]/\gamma,\nonumber\\
\Delta_{\tilde{\mathcal{B}}_{1,2}}&=&\sqrt{2S^2/\bar{\gamma}+\tilde{S}^2}\mp\tilde{S},
\end{eqnarray}
where $\bar{\gamma}\equiv \gamma-1/2$.
\par We can read off the following facts from Fig.~\ref{mainresult}(a):
\par (a) For $\gamma<1/2$, $\mathcal{L}_{\Delta,\gamma}$ does not intersect $\tilde{\mathcal{B}}_{1,2}$. If $(\Delta,d)\in \mathrm{I}$, the Ex BS emerges at $k_1=-2\arccos\Delta/\Delta_{\tilde{\mathcal{C}}_1}$, and a higher-energy SI BS is developed at $k_2=-2\arccos\Delta/\Delta_{\tilde{\mathcal{C}}_2}$. We would like to mention that Tonegawa obtained the same expression for $k_2$ using the Green's function method [see Eq.~(3.9) there]~\cite{Tonegawa1970}. We figuratively express the above flow as
\begin{eqnarray}
\mathrm{no~BS}^{(\mathrm{I})} \stackrel{k_1}{\rightharpoondown}\mathrm{Ex}^{(\mathrm{i})} \stackrel{k_2}{\rightharpoondown}\left(
                                      \begin{array}{c}
                                        \mathrm{SI} \\
                                        \mathrm{Ex} \\
                                      \end{array}
                                    \right)^{(\mathrm{iii})},
\end{eqnarray}
where the SI in the last term is arranged above the Ex to indicate that it has a higher energy. If $(\Delta,d)\in \mathrm{i}$, the Ex BS exists throughout the zone and the SI BS still emerges at the $k_2$. If $(\Delta,d)\in \mathrm{iii}$, then the Ex and SI BSs simultaneously exist throughout the zone.
\par (b) For $1/2<\gamma<1/2+S/[2(S+ \tilde{S})]$, $\mathcal{L}_{\Delta,\gamma}$ may intersect $\tilde{\mathcal{C}}_1$, $\tilde{\mathcal{B}}_1$, $\tilde{\mathcal{C}}_2$, and $\tilde{\mathcal{B}}_2$ at $\Delta_{\tilde{\mathcal{C}}_1}<\Delta_{\tilde{\mathcal{B}}_1}<\Delta_{\tilde{\mathcal{C}}_2}<\Delta_{\tilde{\mathcal{B}}_2}$. If $(\Delta,d)\in \mathrm{I}$, the Ex BS again emerges at $k_1$. When $k$ further decreases to $q_1=-2\arccos\Delta/\Delta_{\tilde{\mathcal{B}}_1}$, the RP enters region ii where the single Ex BS transforms to SI-type. The other higher SI BS shows up in region iv as we lower $k$ down to $k_2$. As $k$ continues decreasing to $q_2=-2\arccos\Delta/\Delta_{\tilde{\mathcal{B}}_2}$, the foregoing higher-lying SI BS becomes an Ex one after entering the final region v. The corresponding flow is expressed as
\begin{eqnarray}\label{flowb}
\mathrm{no~BS}^{(\mathrm{I})} \stackrel{k_1}{\rightharpoondown} \mathrm{Ex}^{(\mathrm{i})} \stackrel{q_1}{\rightharpoondown} \mathrm{SI}^{(\mathrm{ii})} \stackrel{k_2}{\rightharpoondown}\left(
                                      \begin{array}{c}
                                        \mathrm{SI}_2 \\
                                        \mathrm{SI} \\
                                      \end{array}
                                    \right)^{(\mathrm{iv})}
                                   \stackrel{q_2}{\rightharpoondown}\left(
                                      \begin{array}{c}
                                        \mathrm{Ex} \\
                                        \mathrm{SI} \\
                                      \end{array}
                                    \right)^{(\mathrm{v})},\nonumber\\
\end{eqnarray}
where the $\mathrm{SI}_2$ stands for the second higher-energy SI BS in region iv.
Similar analysis can be applied to the cases of $(\Delta,d)\in \mathrm{i},\mathrm{ii},\mathrm{iv}$, or $\mathrm{v}$.
\par (c) For $\gamma>1/2+S/[2(S+ \tilde{S})]$, $\mathcal{L}_{\Delta,\gamma}$ may intersect $\tilde{\mathcal{C}}_1$, $\tilde{\mathcal{B}}_1$, $\tilde{\mathcal{C}}_2$ at $\Delta_{\tilde{\mathcal{C}}_1}<\Delta_{\tilde{\mathcal{B}}_1}<\Delta_{\tilde{\mathcal{C}}_2}$. The flow reads
\begin{eqnarray}
\mathrm{no~BS}^{(\mathrm{I})} \stackrel{k_1}{\rightharpoondown} \mathrm{Ex}^{(\mathrm{i})} \stackrel{q_1}{\rightharpoondown} \mathrm{SI}^{(\mathrm{ii})} \stackrel{k_2}{\rightharpoondown}\left(
                                      \begin{array}{c}
                                        \mathrm{Ex} \\
                                        \mathrm{SI} \\
                                      \end{array}
                                    \right)^{(\mathrm{v})}.
\end{eqnarray}
\par We now apply the above general analysis to the examples in Fig.~\ref{motivation}. The evolution of the RPs for the three examples with $(\Delta,\gamma)=(1,1)$, $(0.25,0.6)$, and $(1.2,0.4)$ are shown in Fig.~\ref{mainresult}(b). The case of $(\Delta,d)=(1,1)\in\mathrm{i}$ belongs to case (c) and the flow goes as $\mathrm{Ex}^{(\mathrm{i})} \stackrel{q_1}{\rightharpoondown} \mathrm{SI}^{(\mathrm{ii})} \stackrel{k_2}{\rightharpoondown}\left(
                                      \begin{array}{c}
                                        \mathrm{Ex} \\
                                        \mathrm{SI} \\
                                      \end{array}
                                    \right)^{(\mathrm{v})}$, with the transition point $q_1=-2\arccos\frac{1}{\sqrt{3}}\approx -3\pi/5$. The higher-energy Ex BS subsequently appears at $k_2=-2\arccos\frac{1}{3+\sqrt{6}}\approx -9\pi/10$. For $(\Delta,d)=(0.25,0.15)\in\mathrm{I}$ we have $\gamma=0.6<3/(6+4\sqrt{3})$, so that it belongs to case (b). The corresponding flow is given by Eq.~(\ref{flowb}). The first Ex BS emerges at $k_1=-2\arccos\frac{4+\sqrt{11}}{20}\approx -19\pi/25$ and transforms to the SI BS at $q_1=-2\arccos\frac{\sqrt{3}}{36}\approx -24\pi/25$. The RP enters region iv at $k_2=-2\arccos\frac{4-\sqrt{11}}{20}\approx -\pi+4\pi/N$, which explains the simultaneous appearance of the two SI BSs in this mode. The case of $(\Delta,d)=(1.2,0.48)\in\mathrm{i}$ can be similarly analyzed.
\par For $k\in K'_1$, the associated transcendental equation has a different form from Eq.~(\ref{tan1})~\cite{SM}. However, the resulting phase diagram is very close to that of $k\in K_1$ and the two tend to be identical as $N\to\infty$~\cite{SM}. We thus achieved a complete solution of the problem.
\par \emph{Conclusions}. We develop a theoretical method for semianalytically solving the two-magnon problem in a higher-spin XXZ chain with single-ion anisotropy. Although several typical features of the evolution of the two-magnon bound states with varying wave number have been observed in prior literature, the understanding of them is limited to qualitative explanations. By combining a set of recently constructed exact two-magnon Bloch states for higher-spin chains and a plane-wave ansatz, we are able to treat the problem as analytical as possible. Explicitly, we establish a complete phase diagram consisting of various parameter regions that support different types of bound states, with the phase boundaries defined by algebraic equations. In particular, we discover a narrow parameter region in which both of the two lowest-lying bound states are of single-ion type. By showing that the phase diagrams for any two different wave numbers are similar to each other, we convert the evolution of the bound states to the movement of a representative point for given parameters on a rescaled phase diagram, which offers quantitative explanations to the previously observed behaviors of the bound states.
\par Considering the spin-1 Heisenberg model with single-anisotropy has recently been realized with ultracold atoms~\cite{Ketterle2021}, our results might be relevant to several timely research directions such as bound-state dynamics in engineered quantum spin systems~\cite{Roos2023,Kim2024} and exciton fission~\cite{Teichen2015} where multilevel structure is relevant. Our method can be applied to more general nearest-neighbor periodic quantum chains, e.g., spin chains including higher-order terms~\cite{Bonner1988} and itinerant particle systems~\cite{Pan2019,Mishra2022}. In addition, two-magnon excitations upon antiferromagnetic states can also be treated by the present approach.
\par \emph{Acknowledgments}. N.W. thanks H. Katsura for suggesting the plane-wave ansatz solution of the problem. This work was supported by the National Key Research and Development Program of China under Grant No. 2021YFA1400803.


\begin{thebibliography}{99}
\bibitem{whylargeN} In principle, one can insert the solutions of $R=\pm 1$ into Eq.~(\ref{tanh1_even}) to get the expressions for $\mathcal{B}_{1,2}$. However, in the resulting equation $d$ connects to $\Delta$ only implicitly.
\end{thebibliography}
\end{document}